\documentclass{emulateapj}
 \shorttitle{Deep Chandra pointing on M84}
 \shortauthors{Finoguenov et al.}

\begin{document}

\submitted{ApJ in press, November 1, 2008, v687n1}

\title{In-depth {\it Chandra} study of the AGN feedback in Virgo elliptical
  galaxy M84}

\author{A. Finoguenov\altaffilmark{1,2}, M. Ruszkowski\altaffilmark{3,6},
  C. Jones\altaffilmark{4},  M. Br{\"u}ggen\altaffilmark{5},
  A. Vikhlinin\altaffilmark{4}, E. Mandel\altaffilmark{4}}

\altaffiltext{1}{Max-Planck-Institut f\"ur Extraterrestrische Physik,
             Giessenbachstra\ss e, 85748 Garching, Germany}
\altaffiltext{2}{University of Maryland, Baltimore County, 1000
  Hilltop Circle,  Baltimore, MD 21250, USA}
\altaffiltext{3}{Department of Astronomy, University of Michigan, 500 Church
  Street, Ann Arbor, MI 48109-1042, USA}
\altaffiltext{4}{Harvard-Smithsonian Center for Astrophysics, 60 Garden Street,
 Cambridge, MA 02138, USA}
\altaffiltext{5}{Jacobs University Bremen, Campus Ring 1, 28759 Bremen, Germany}
\altaffiltext{6}{Max-Planck-Institut f\"ur Astrophysik, D-85748 Garching, Germany}

\begin{abstract}

  Using deep {\it Chandra} observations of M84 we study the energetics of
  the interaction between the black hole and the interstellar medium of this
  early-type galaxy. We perform a detailed two dimensional reconstruction of
  the properties of the X-ray emitting gas using a constrained Voronoi
  tessellation method, identifying the mean trends and carrying out the
  fluctuation analysis of the thermodynamical properties of the hot ISM.  In
  addition to the $PV$ work associated with the bubble expansion, we
  identify and measure the wave energy associated with the mildly supersonic
  bubble expansion. We show that, depending on the age of the cavity and the
  associated wave, the waves can have a substantial contribution to the
  total energy release from the AGN. The energy dissipated in the waves
  tends to be concentrated near the center of M84 and in the direction
  perpendicular to the bubble outflow, possibly due to the interference of
  the waves generated by the expansion of northern and southern bubbles.  We
  also find direct evidence for the escape of radio plasma from the ISM of
  the host galaxy into the intergalactic medium.

\end{abstract}

\keywords{AGN -- galaxies: intergalactic medium -- galaxies: elliptical --
  X-rays: galaxies}

\section{Introduction}

The detailed investigations of the balance between the heating and cooling
of the hot interstellar medium (ISM) and intracluster medium (ICM) have been
made possible due to the key observations made by {\it Chandra} and XMM-{\it
  Newton} and the successes in numerical modeling (see McNamara \& Nulsen
2007 for a review).  The commonly accepted paradigm states that the feedback
mechanism is ultimately linked to the activity of the supermassive black
holes located
at the bottom of the potential wells in systems with cool cores.\\
\indent A standard way of estimating the energy released by the AGN is to
measure the $PV$ work associated with X-ray cavities (bubbles) inflated by
the AGN.  This is an indirect measurement as the bubbles are typically
``observed'' as depressions in X-ray emissivity.  However, the task of
estimating the AGN energy injected into the ICM can be accomplished by
assuming that such X-ray cavities are in pressure balance with the ICM. One
can then infer the total energy contained in the cavities that is available
for doing mechanical work (i.e., enthalpy) from $H=\gamma/(\gamma -1)PV$,
where $\gamma$ is the adiabatic index, $P$ is the pressure and $V$ is the
bubble volume.  Such an observational estimate is further complicated by the
fact that the effective adiabatic index $\gamma$ of the material inside the
bubbles is not known.  That is, it is not known if the bubbles are
predominantly thermal, in which case $\gamma = 5/3$ (e.g., Mazzotta et al.
2002 shows that both the thermal and non-thermal models fit the data in the
case of MKW 3s galaxy cluster) or non-thermal, in which case $\gamma = 4/3$
(e.g., Sanders \& Fabian 2007 in the case of the Perseus cluster). The
content of radio lobes and X-ray cavities may depend on such factors as the
initial jet composition ($e^{+}-e^{-}$ or $p-e^{-}$), the efficacy of
entrainment of colder thermal ICM gas or the magnetic pressure support
inside the bubbles (Dunn et al. 2006 and references therein).  Furthermore,
the energy content obtained this way may be a lower limit to the actual
total energy released by the AGN if the pressure balance assumption does not
hold.  An extreme example of this effect is seen in the simulations of very
high Mach number jets interacting
with the ICM (Binney, Bibi \& Omma 2007). \\
\indent Moreover, X-ray observations show that the standard method of
inferring AGN energies leads to the bubble energy {\it increasing} in the
statistical sense with the distance from the center of the gravitational
potential (Diehl et al. 2008).  This apparently counter intuitive result may
be the result of the erroneous assumption that the bubbles are ``born'' in
pressure
equilibrium with the surrounding ICM or ISM.\\
\indent If the bubbles are overpressured with respect to the ISM, then one
would need the internal bubble pressure to correctly estimate the energy
inside the bubble.  Moreover, if they were overpressured in the past, then
this overpressure should have created waves that have carried some portion
of the energy away from the cavities. The standard method of estimating the
AGN energy based on the pressure balance and cavity size does not include
such contributions to the total energy balance.  In addition to increasing
the overall energy budget, such waves provide a more gentle and
spatially-distributed heating.\\
\indent The above effects clearly demonstrate the need for alternative or
supplementary measurements of the energy injected by the AGN. In this Paper
we report on the long time exposure observation of the elliptical galaxy M84
in the Virgo cluster. We focus on the interaction of the AGN with the ISM
and the energy content in the thermodynamical fluctuations generated by the
the AGN outburst and argue that they are waves. An approach to measure the
wave energy that is similar to the one described in this Paper, has been
considered by Sanders \& Fabian (2007) in the case of the observations of
the Perseus cluster and in the numerical simulations of wave dissipation
(Ruszkowski et al. 2004).

\section{Chandra observations}

Results from the first Chandra observation of M84 (OBS ID 803 and exposure
time of 25.9 ksec) were published in Finoguenov \& Jones (2001, 2002).
Given the importance of the source in understanding the AGN feedback, we
were granted two more observations (OBS IDs 5908 and 6131) adding 34.9 and 30.9
ksec time, respectively. The nominal aim point was ACIS-S and we only
include S3 CCD data here. The initial data reduction is standard and its
details are presented in Vikhlinin et al. (2005).

\includegraphics[width=8.0cm]{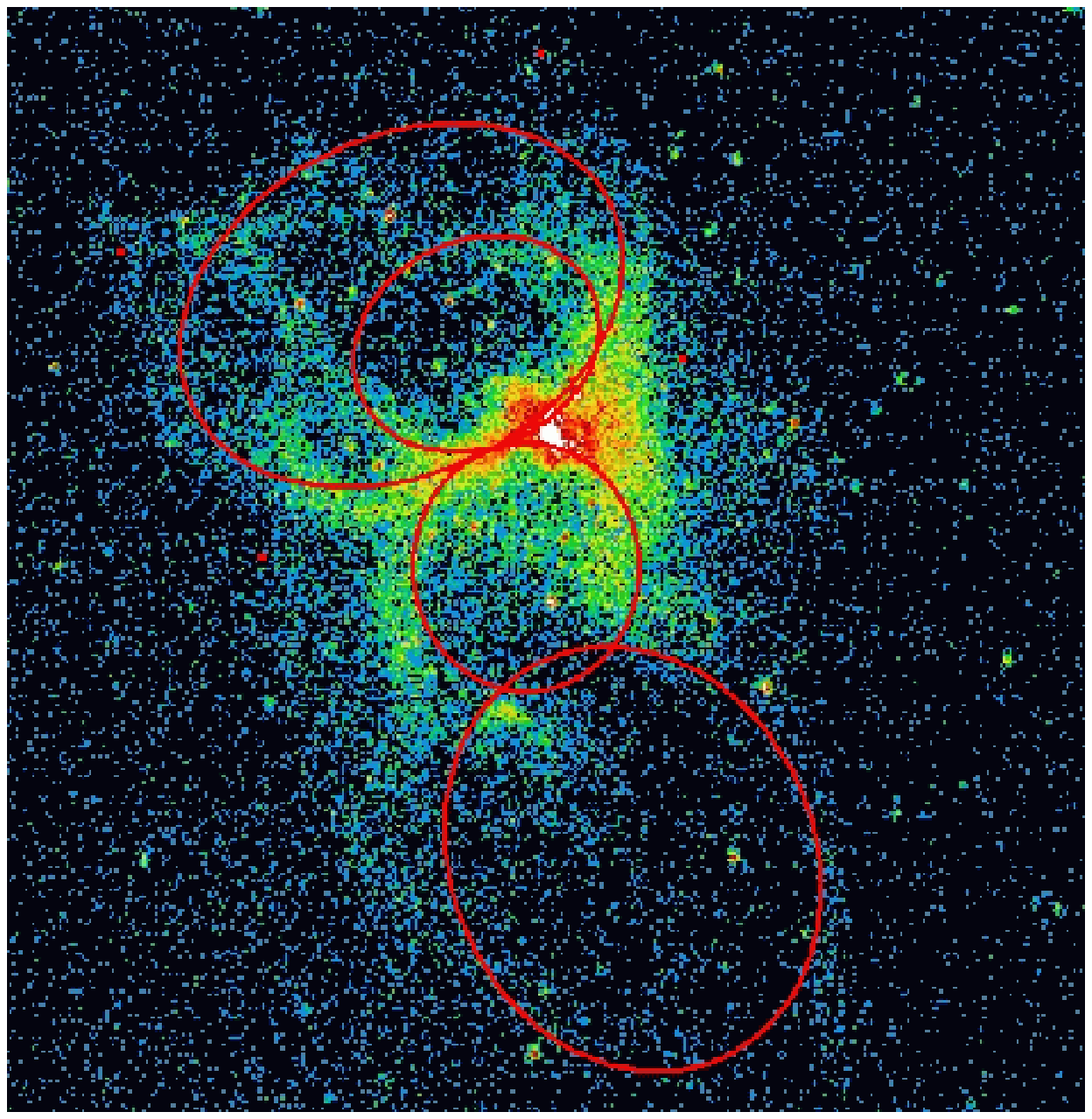}
\figcaption{Image of M84 in the 0.5--2 keV band. Red circles show the
  position and size of the four identified bubbles. The image is 2.8
  arcminute or 14 kpc wide on a side.}

For both imaging and spectroscopic analysis we
extracted the counts and the auxiliary information separately from each
observation, and added it together at a final stage proportional to the
exposure time. The distant-dependent parameters of the emission have been
calculated assuming a 17 Mpc distance to M84 at which 1 arcminute
corresponds to 5.0 kpc.

\subsection{Imaging}

The raw cumulative photon count ACIS-S image of M84 in the energy band 0.5--2
keV is shown in Fig.1. In addition to previously identified cavities
associated with the radio bubbles, we clearly see structure in the bubbles,
which can be approximated by two sets of two bubbles to the north and south
of the M84 center.  In Fig.2 we compare the X-ray and radio properties of
M84. It is clearly seen that while the southern radio bubble is surrounded
by the X-ray cocoon, the northern radio bubble broke through the porous X-ray
emission at the northern edge. This provides direct evidence of escaping
radio plasma from the ISM of the host galaxy into the intergalactic
medium.

This paper describes the parameters of these bubbles, with a summary of the
results given in Table 1, listing name of the bubble (col.  1), position of
the bubble centroid (2-3), bubble axes (major, minor, assumed projected) in
kpc (4-6), effective thickness of bubble walls (7), total geometrical factor
from Eq.2 (8), estimated Mach number from the pressure jump (9), $PV$ work
(enthalpy $H$ is higher by up to a factor of 4, as explained above) (10),
wave energy, calculated using Eq.2 with $\gamma=5/3$ (11; see below). In
calculating the values of columns 10 and 11 we used a 3-d integral over the
bubbles and the non-parametric fit to the pressure profile of all zones of
M84, as explained in \S 3. In calculation we use the exact pressure prediction
at each dV element assuming a spherical symmetry in the pressure.

Imaging analysis provides information on the bubble appearance, such as
centering, size and orientation, which enter the volume and surface
calculations. The width of the bubble walls is calculated as the ratio between
the distances from the bubble center to the outer and the inner boundaries,
which is used in the calculation of the wave energetics; and helps to select
the regions for subsequent spectral analysis, which we describe next.

\includegraphics[width=8cm]{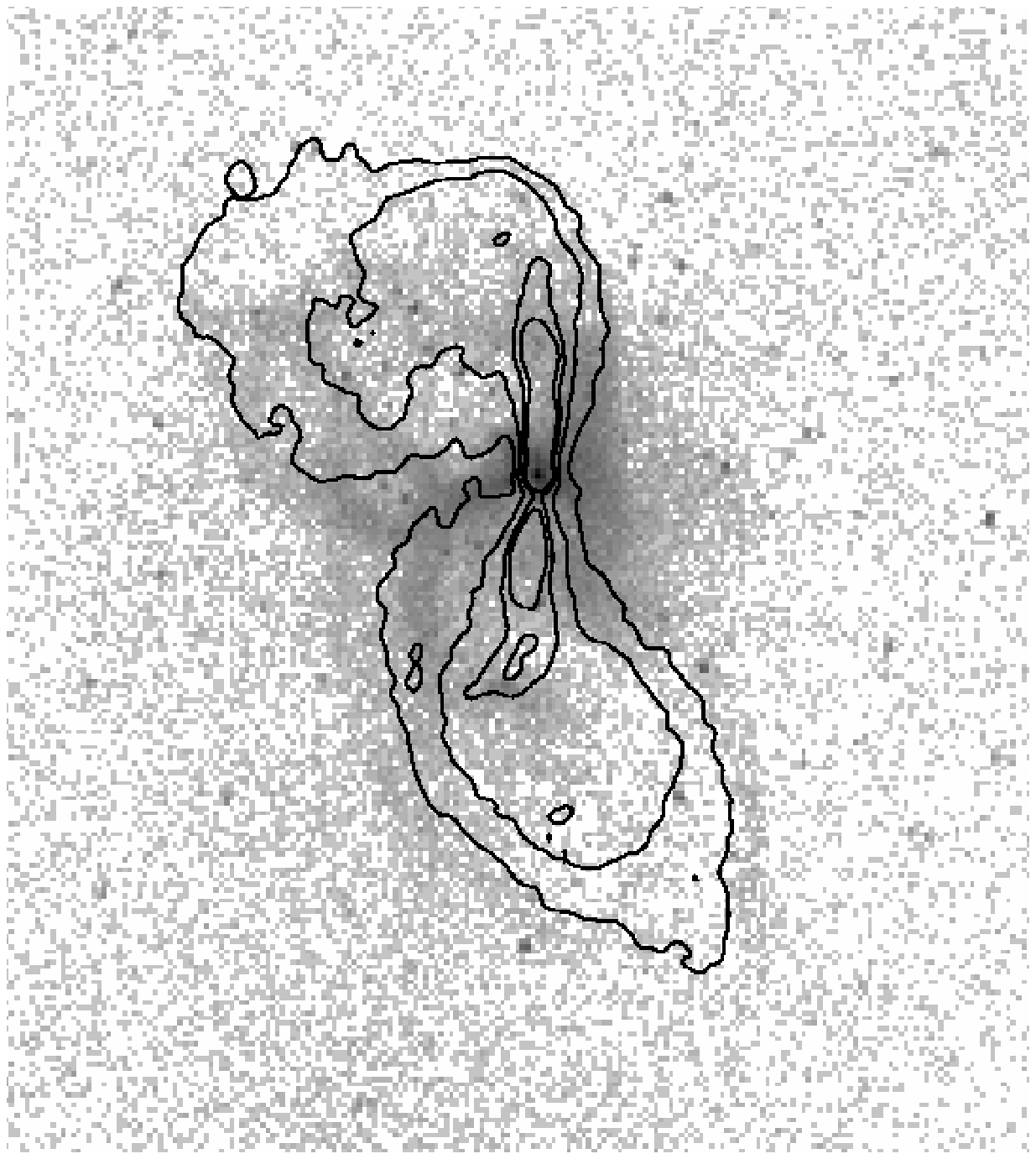} 
\figcaption{Binned raw X-ray image of M84 with radio contours
  overlaid. For the southern part of M84, radio emission is clearly embedded
  in the X-rays, while in the north it breaks through the
  bubbles. The image is 3.6
  arcminute or 18 kpc wide on a side. }

Statistics, achieved in the combined {\it Chandra} observation of M84,
requires substantial binning of the data for the subsequent spectral
analysis. The binning techniques based on the Voronoi tessellation methods
have been proven to be the most efficient and unbiased way to address the
binning issue (e.g. Cappellari \& Copin 2003). However, in order to study the
waves and bubbles, we need to separate the bubble rims from both the inner
and outer medium.  To accomplish this task, we produced masks defining large
contiguous zones of equal emissivity, which are then subsequently subdivided
in order to achieve the selected signal-to-noise ratio using the Voronoi
tessellation.  Previous applications of this technique is given by
Simionescu et al. (2007) and is somewhat similar to the contour binning
technique of Sanders et al.  (2006).

\section{Maps}
Using the constraints imposed by strong variations in the X-ray surface
brightness, we have designed a mask of 130 regions which depict all the
important details of M84.  This mask optimally splits large regions into
smaller ones using the Voronoi tessellation method. This method has been
previously applied to XMM-{\it Newton} observations of M87 (Simionescu et al.
2007) and A3128 (Werner et al. 2007). This is a first application of this
method to {\it Chandra} data. We published the results of this analysis
using the meta-table format developed within the German Astrophysical
Virtual Observatory (GAVO) and published as a web-service accessible at {\it
  http://www.g-vo.org/MAXI} and having a data release unique identifier
(druid) of {\it http://www.g-vo.org/MAXI/druid/2}.
This service allows one to view the parameters of the fit, access the
quality of the fit, build, view and retrieve any maps and access the
significance of every feature.

\begin{figure*}
\includegraphics[width=18cm]{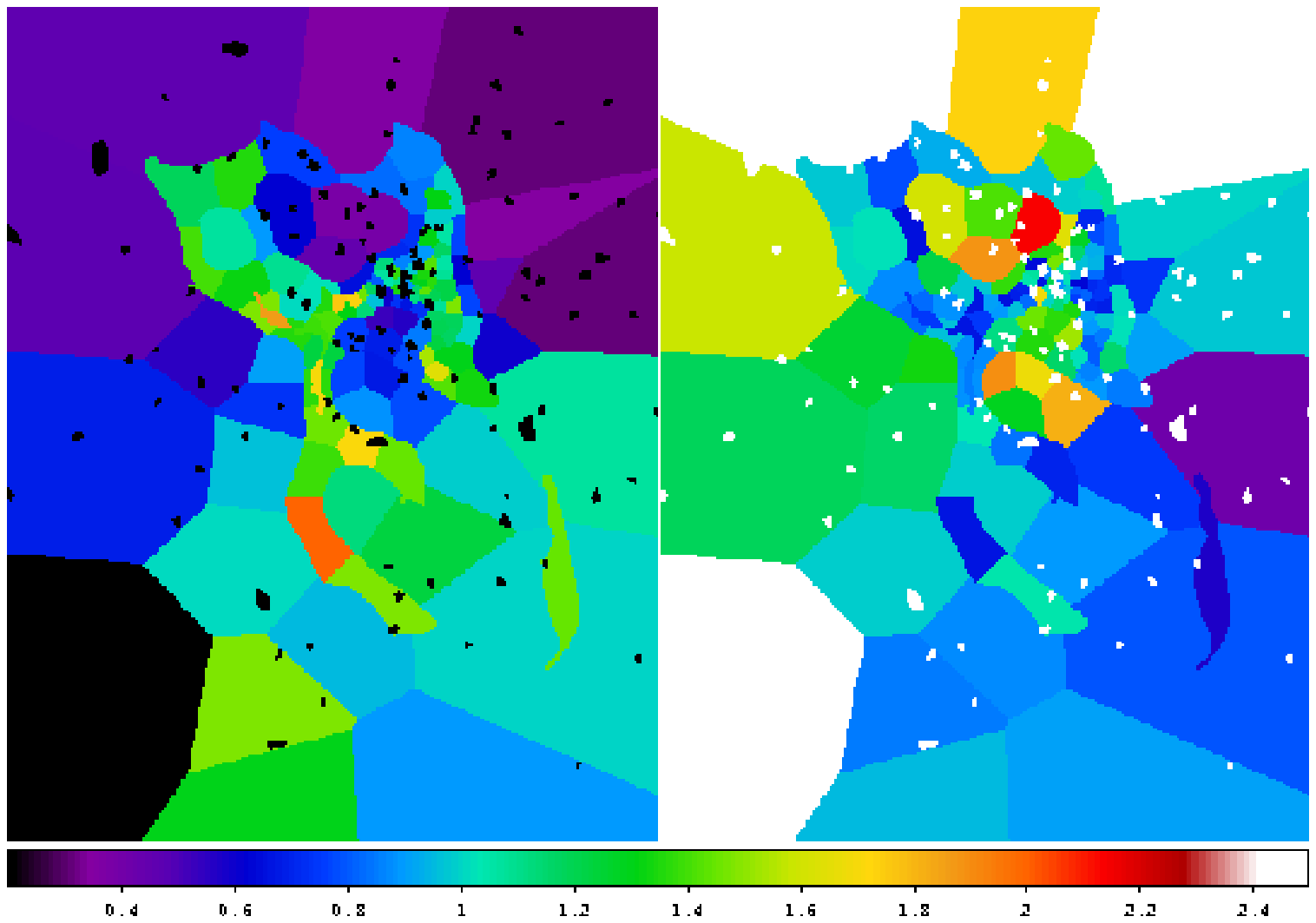}
\figcaption{ Ratio of the observed pressure {\it left} and
  entropy {\it right} in M84 to their corresponding 
  average profiles. The values vary between 0.5 and 2. One can identify the
  bubble walls with a factor of 1.5 higher pressure and 40\% lower
  entropy. Each image is $2.8\times3.8$ arcminutes ($14\times19$ kpc) in size.}
\end{figure*}

In the spectral analysis, we used the APEC model. In addition to the soft
emission of M84,
we clearly detect harder emission, which on smaller scales is centered on
M84 and has been discussed in Finoguenov \& Jones (2001), while on large
scales is associated with emission of the Virgo cluster in which M84 is
embedded. Thus, we introduced a second APEC model with temperature fixed at
3 keV and a fixed metalicity of 0.3 solar. Having two thermal models
substantially reduces our ability to derive the metalicity in M84.
Therefore, we have also fixed the metalicity of the APEC component
describing the M84 emission to 0.3 solar, which is typical of the regions of
M84 with bright X-ray emission.

Since the bubbles are symmetric about their center, but not with respect to
the center of X-ray emission in M84, we have adopted a complex procedure for
deriving the volumes, which require an estimate of the projected length. All
bubble-related features are calculated assuming spherical symmetry relative
to bubble centers, reported in Table 1. Other features are calculated using
spherical symmetry relative to the center of X-ray brightness of M84. The
exact details of our volume calculation and selecting the centers of the
elements are presented in Mahdavi et al. (2005).

In order to study the fluctuations in the pressure and entropy, first we
have analyzed the mean trends, following the procedure outlined in Sanderson
et al. (2005), which puts the values of the map on the profile according to
the distance to the center of the region, generates a non-parametric fit to
the profile (using {\it R}-package) and calculates the residuals. Each
region is treated as one point on the profile. Within this procedure the
small differences in defining the center of the region (as e.g. discussed in
Mahdavi et al. 2005) would result in differences in the mean profile but not
in the ratios. In Figure 3 we show the ratio of the observed pressure to the
mean pressure profile (left panel) and the analogous quantity for the gas
entropy.  The fractional rms fluctuations of entropy and pressure caused by
the AGN are on the level of 47\% and 41\%, respectively, with 5\%
measurement uncertainty.  For comparison, in clusters of galaxies similar
levels of fluctuations are associated with distortions due to a recent
merger (Poole et al. 2007) and has about 10\% occurance probability at low
redshift (Finoguenov et al.  2005, 2007).  However, the features associated
with cluster mergers appear on much larger spatial scales.

\section{Hydrodynamical simulations}

In order to gain insights into the bubble physics, we embarked on 
hydrodynamical simulations. The details of the simulation used here can be
found in Br{\"u}ggen, Ruszkowski \& Hallman (2005). Here we only summarize the
relevant information. The initial conditions for the simulation were taken
from the S2 cluster run (Springel et al. 2001) performed with the {\it
  GADGET} code. Starting from these initial conditions (at $z=0$) we evolved
the system for 140 Myr using the adaptive mesh refinement {\it FLASH} code.
The full size of the computational domain was 2$h^{-1}$ Mpc and the maximum
resolution was 1.96$h^{-1}$ kpc. While the cluster atmosphere and ICM
parameters differ from those in M84, the initial conditions possess some
characteristics that make the simulated system are qualitatively similar to
M84. In particular, the structure is quite dynamic, the central object moves
relative to the surrounding gas and the temperature in the central parts
raises with radius as in M84.

\section{Waves}

The main result from the maps in Fig.3 is the identification of the bubble
walls with the regions of enhanced pressure (by a factor of 1.5) and
simultaneously 40\% lower entropy. Previously, the presence of shock waves
has been often dismissed on grounds of low temperature contrast seen between
the bubbles and the surrounding medium. Such comparison assumed a similarity
in the entropy between the bubble walls and the surrounding medium. This is
in contradiction to the observed entropy distribution seen in Figure 3,
which shows that the gas associated with the bubble walls has low entropy.
One would expect higher temperature in the compressed gas, if it were not
for the fact that low temperature gas is entrained and moved to larger
distances from the center thus partially cancelling the effect of adiabatic
heating.  Thus low temperature contrast may well be consistent with the wave
scenario.  We estimated that, on the time-scale of the bubble expansion
($10^7$ yr), the cooling (time scale of $10^9$ yr) is not important and the
origin of low entropy gas is due to entrainment from the central regions of
M84. The entropy profile in M84 is rather steep and small gas displacements
(compared to the bubble size) are sufficient to reproduce an observed
picture.

\begin{deluxetable*}{lccccccccc}
\centering
\tablewidth{0pt}
\tabletypesize{\footnotesize}
\tablecaption{Characteristics of bubbles in M84}
\tablehead{
\colhead{ } &
\colhead{ position of the center} &
\colhead{ axes} &
\colhead{ $\lambda/r$} &
\colhead{$(\lambda/r)\times$} &
\colhead{ } &
\colhead{ $PV$ work} &
\colhead{ wave energy}\\
\colhead{Bubble name } &
\colhead{ RA, Dec. (Eq. 2000)} &
\colhead{ kpc} &
\colhead{ } &
\colhead{$(r/R)^3$} &
\colhead{ $M$} &
\colhead{  $10^{55}$ ergs} &
\colhead{  $10^{55}$ ergs }}

\startdata
Southern Large &186.26217 +12.869655& 2.6 2.2 2.2& 1/7 &0.2 &1.3 & 1.32 & 0.67 \\
Southern Small &186.26680 +12.882228& 1.4 1.5 1.4& 1/4 &0.6 &1.3 & 0.30 & 0.37 \\
Northern Small &186.26876 +12.890701& 1.3 1.5 1.5& 1/4 &0.6 &1.3 & 0.26 & 0.35 \\
Northern Large &186.27185 +12.892204& 2.0 2.9 2.9& 1/7 &0.2 &1.3 & 1.37 & 0.72 \\
\enddata

\end{deluxetable*}

\indent We thus interpret the overpressured shells seen in M84 as waves
propagating away from the sites where the energy has been injected by the
AGN.  The compression of the gas immediately outside the cavity is released
in the form of a weak shock wave. Similar interpretation has also been
presented in case of Perseus cluster (Fabian et al. 2003) and M87 (Forman et
al. 2007). This interpretation is supported by the results of numerical
simulations (see Fig.4). We also note that the observed fractional pressure
fluctuations in the ``walls'' are larger than the fractional density
fluctuations, which is consistent with the adiabatic compression scenario.
As a proof of concept, in Figure 4 we show a density slice corresponding to
a snapshot from an adaptive mesh refinement hydro simulation of AGN
feedback.  The images have been unsharp-masked and Gaussian-blurred to
enhance the fluctuations.  There is a qualitative similarity between this
figure and the morphology of M84.  This figure shows that multiple outbursts
can lead to nested waves similar to a Russian matryoshka doll.  The upper
panel corresponds to an earlier epoch and the snapshots are separated by
approximately $2\times 10^7$ years. The waves detach from the AGN-inflated
cavities as is clearly seen in the lower panel.  The northern cavity is
further distorted due to the relative motion between the AGN and the ICM.
Interestingly, the evolution of the (nearly) vertical boundary between the
northern bubbles seen in the simulation, shows that this feature is also a
propagating wave (this is clearly seen in the animated version of the data).
A similar feature is located "inside" the northern cavity in M84.\\

\section{Estimating AGN work}
We estimate the energy carried in the waves using the following approach.
The instantaneous energy flux $F$ carried by a wave is given by:

\begin{equation}
F\equiv\frac{P_{\rm wave}}{S}=\frac{(\delta P)^{2}}{\rho\upsilon_{\rm wave}}, 
\end{equation}

\noindent
where $\upsilon_{\rm wave}$ is the wave propagation velocity, $P_{\rm wave}$
is the wave power, $S=4\pi r^2$ where $r$ is the distance of the wavefront
from the bubble center, $\rho$ is the ICM density and $\delta P$ is the
pressure fluctuation in the wavefront (Landau \& Lifshitz 1987, Ruszkowski,
Br{\"u}ggen, \& Begelman 2004, Sanders \& Fabian 2007).  Strictly speaking,
Equation 1 is valid only for small perturbations, but it should suffice for
our estimates as the inferred Mach number of the waves only slightly exceeds
unity (see below).  We assume that at any given time the wave front is a
sphere, the instantaneous power of the entire wave is: $P_{\rm wave} = 4\pi
r^{2}F \;\; [{\rm erg}\; {\rm s}^{-1}]$.  In the absence of dissipation,
this (total) power would remain constant even though $r$, $\rho$,
$\upsilon_{\rm wave}$ and $\delta P$ would all vary.  The total energy
carried by one wave is then $E_{\rm wave} \sim P_{\rm
  wave}\lambda/\upsilon_{\rm wave}$, where $\lambda$ is the wavelength (or
$\sim$ the thickness of the pressure fluctuation). Approximating
$\upsilon_{\rm wave}$ as the adiabatic sound speed, the final expression for
the wave energy is:

\begin{equation}
E_{\rm wave}\sim 3 \left (\frac{\lambda}{r}\right )\gamma^{-1}PV\left (\frac{\delta P}{P}\right )^{2}\left (\frac{r}{R}\right )^{3},
\end{equation}

\noindent
where $V$ is the bubble volume, $P$ is an underlying pressure, $R$ is the
bubble radius, $r$ is the distance of the wavefront from the bubble center
and $\gamma$ is the adiabatic index of the gas in the vicinity of the wave.
The pressure morphology of the source suggests that the ratio of the wave
thicknesses to the radii of the wavefronts is approximately in the range
$\sim 1/7$ to $1/4$ depending on the bubble position (e.g., $\sim 1/4$ for
the small bubbles, $\sim 1/7$ for the large bubbles). If $r\sim R$, and
using our observation of $\delta P/P\sim 1$, this would suggest that $E_{\rm
  wave}$ is of the order of $PV$.  Even if the bubble energy is greater than
$PV$ by a few, one wave carries a significant fraction of the outburst
energy.  The result of this analysis is shown in Table 1, where wave energy
is compared to $PV$ work. The energy carried by the waves can significantly
contribute to the overall energy budget.  Note also that the smaller inner
outbursts have larger wave-to-bubble energy ratios.


In addition to energy carried by the wave, we can calculate the energy
deposition into the IGM associated with supersonic wave motion, 

\begin{equation}
  dQ = T dS =
  {1 \over \gamma - 1 } \left ({kT \over \mu m_p} \right ) {dS_x \over S_x}.
\end{equation}

where $S_x=kT \rho^{1-\gamma}$ is an common definition of the entropy among
X-ray observers.  For the value of Mach number of 1.3 measured for M84
bubbles, and employing the Rankine-Hugoniot adiabat shock adiabat to derive
the change in the entropy

\begin{equation}
1+ {dS_x \over S_x} = {(2 \gamma M^2 - (\gamma-1))((\gamma-1) M^2+2)^\gamma \over (\gamma+1)^{\gamma+1} M^{2\gamma}}
\end{equation}

${dS_x \over S_x}=0.013$ and the full energy deposition $dQ dM_{gas}$ is a
few percent of the total wave energy.  The maximum deposition of energy
occurs near the center of M84 and in the direction perpendicular to the
bubble outflow, where also the cooling looses are largest. This probably
results from interference of waves coming from northern and southern bubbles
and is an important channel of energy deposition into the central parts of
M84 IGM.

We note that using
\begin{equation}
 {dS_x \over S_x} = {dP \over P} - \gamma {d \rho \over \rho}
\end{equation}

 $dQ dM_{gas}$  can be
rewritten as

\begin{equation}
dQ dM_{gas}=\int { dP \over \gamma -1} dV - \int  {\gamma \over \gamma -1} P
{ d \rho \over \rho} dV
\end{equation}

The second part of the equation is negligible in the case of strong shock
($M>1000$) which leads to the equation used by Graham et al. (2008). We
note, however, that in our case the two terms are nearly equal and in case
of adiabatic compression they exactly cancel out.  The total heating
supplied by AGN, when compared to $9.4\times 10^{40}$ bolometric luminosity
associated with {\it extended} emission of M84 (not counting the
contributions from AGN and LMXB), requires a bubble recurrence on the
$1.8\times 10^7$ yr scale to entirely compensate for the radiative energy
losses, similar to what is actually observed in M84.

\hspace*{-1cm}\scalebox{0.8}{\includegraphics[]{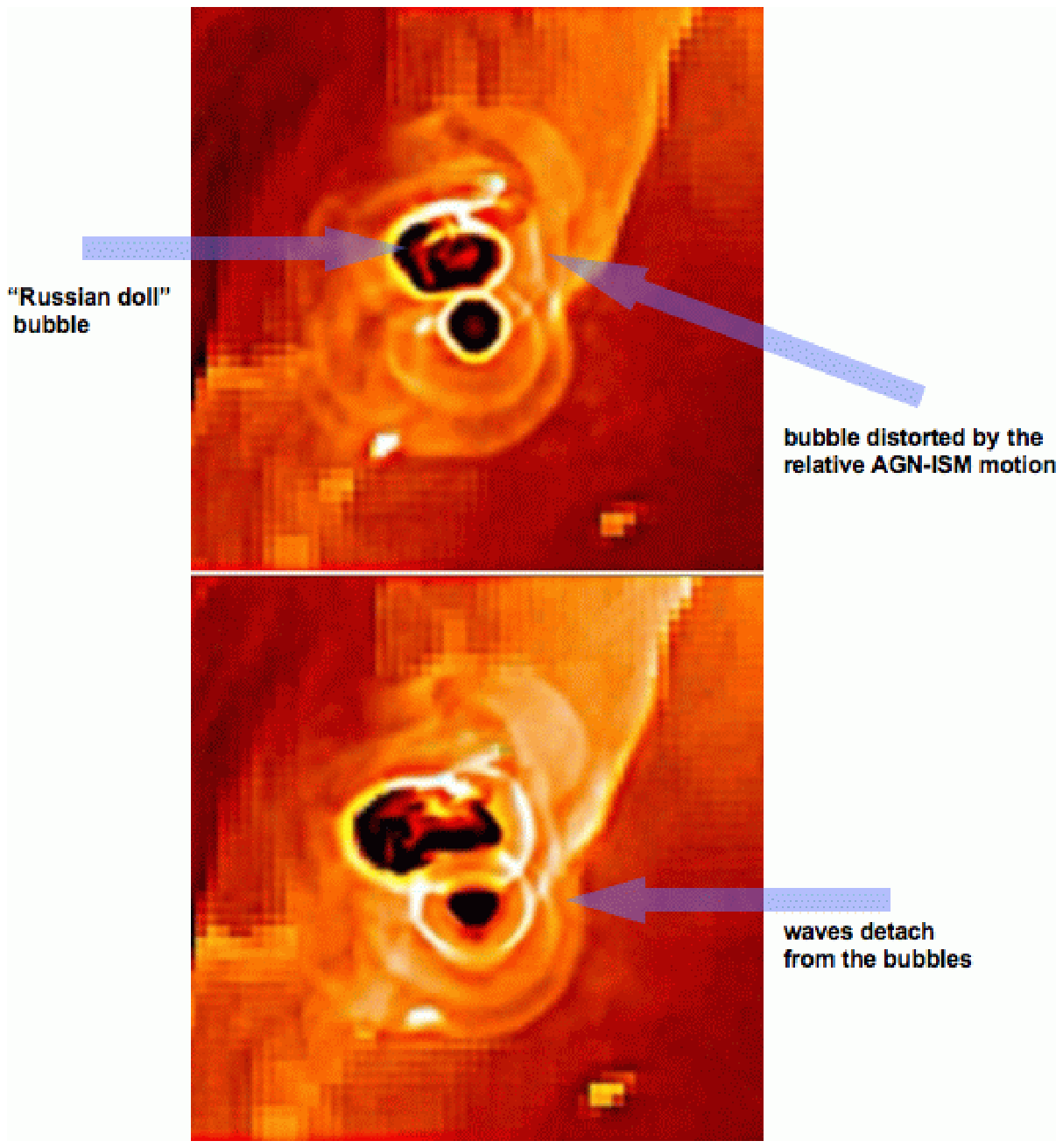}}
\figcaption{Subsequent density snapshots from an adaptive mesh simulation of
  AGN feedback in a galaxy cluster. The box sizes in both panels are about 300$^{-1}h$ kpc on a side. See text for details.}

\indent The morphology of the southern shells seen in the data suggests that
the expansion velocity is comparable to the velocity of M84 relative to the
ICM.  A typical velocity of a cluster galaxy is expected to be mildly
supersonic (Faltenbacher et al. 2006).  This is because $c_{s}^{2}=\gamma
P_{\rm gas}/\rho =\gamma \sigma_{\rm gas, 1D}^{2}=\gamma\sigma_{\rm gas,
  3D}^{2}/3= \gamma\sigma_{\rm gal}^{2}/3$, where $\sigma_{\rm gas, 1D}$ and
$\sigma_{\rm gas, 3D}$ are the one-dimensional and three-dimensional gas
internal velocity dispersions, respectively, and $\sigma_{\rm gal}$ is the
galaxy velocity dispersion (Sarazin 1988). The last approximate equality
comes from the assumption of equipartition between the gas and the galaxies.
For $\gamma =5/3$ this leads to the galaxy Mach number $\sim 1.34$.  This in
turn suggests that the shell expansion may be, at least initially,
supersonic. If so, this would also be consistent with an estimate of the
Mach number of the waves.  If the pressure fluctuation is $x=\delta P/P\sim
1$ as suggested by our observations, then for $\gamma =5/3$, the Mach number
of the waves is $M\sim [(1+\gamma^{-1})x/2 + 1]^{1/2}\sim 1.34$ in
qualitative agreement with the estimate above.

\section{Conclusions} 

We have analyzed a deep {\it Chandra} observation of the AGN feedback in the
Virgo elliptical galaxy M84. We have applied a constrained Voronoi
tessellation binning method to {\it Chandra} data. Our main results are: (1)
the AGN outflow is mildly supersonic, (2) the non-thermal plasma from the
AGN-inflated cavities appears to be porous and the non-thermal particles
escape from the cavities, (3) co-centric density perturbations (weak shock
waves) are present and their morphology is qualitatively consistent with the
results of numerical simulations, (4) we have estimated the mechanical
energy in the waves and found that it may contribute substantially to the
overall mechanical energy delivered by the AGN.

\acknowledgments

AF acknowledges support from BMBF/DLR under grant 50 OR 0507. AF thanks CfA
for hospitality during his visits. AF thanks Jaiwon Kim and Gerard Lemson
for their help in publishing the data through GAVO. MR acknowledges {\it
  Chandra} Theory grant TM8-9011X. MB acknowledges the support by the DFG
grant BR 2026/3 and the supercomputing grants NIC 2195 and 2256 at the John
Neumann Institut (NIC) in Julich. The results presented were produced using
the {\it FLASH} code, a product of the DOE ASC/Alliances-funded Center for
Astrophysical Thermonuclear Flashes at the University of Chicago. The
authors thank the annonymous referee for insightful report, that helped to
improve the material presented in this paper.


 \end{document}